\documentclass[aps, twocolumn, showpacs,
preprintnumbers,superscriptaddress]{revtex4}%
\usepackage{amsfonts}
\usepackage{amsmath}
\usepackage{amssymb}
\usepackage{graphicx}%
\begin{document}
\title{Pattern formation of indirect excitons in coupled quantum wells}
\author{C. S. Liu}
\affiliation{Institute of Theoretical Physics and Interdisciplinary Center of Theoretical
Studies, Chinese Academy of Sciences, P. O. Box 2735, Beijing 100080, China}
\affiliation{Department of Physics, National Taiwan Normal University, Taipei 11650, Taiwan}
\author{H. G. Luo}
\affiliation{Institute of Theoretical Physics and Interdisciplinary Center of Theoretical
Studies, Chinese Academy of Sciences, P. O. Box 2735, Beijing 100080, China}
\author{W. C. Wu}
\affiliation{Department of Physics, National Taiwan Normal University, Taipei 11650, Taiwan}
\date{\today}

\begin{abstract}
Using a nonlinear Schr\"odinger equation including short-range two-body
attraction and three-body repulsion, we investigate the spatial distribution
of indirect excitons in semiconductor coupled quantum wells. The results
obtained can interpret the experimental phenomenon that annular exciton cloud
first contracts then expands when the number of confined excitons is increased
in impurity potential well, as observed by Lai \emph{et al.} [Lai $et~al.$,
Science \textbf{303}, 503 (2004)]. In particular, the model reconciles the
patterns of exciton rings reported by Butov \emph{et al.} [Butov $et~al.$,
Nature \textbf{418}, 751 (2002)]. At higher densities, the model predicts much
richer patterns, which could be tested by future experiments.

\end{abstract}

\maketitle

\section{Introduction}

Recently great progress has been made on the research of exciton Bose-Einstein
condensation (BEC) by making use of indirect excitons (spatially separated
electron-hole pairs) in coupled quantum wells (CQW). The advantage of the
indirect excitons is that they have a long lifetime and a high cooling rate.
With these merits, Butov \textit{et al.} have successfully cooled the trapped
excitons to the order of 1K. \cite{Butov2002a} Although there are not enough
evidences to prove that the excitons are in BEC state, it is interesting to
observe several interesting photoluminescence (PL) patterns.\cite{Butov2002b}
Independently Snoke \emph{et al.} have also observed somewhat analogous PL
patterns in a similar system.\cite{Snoke2002a} The key features in the
experiment of Ref.~\cite{Butov2002b} are as follows. (i) Two exciton rings are
formed. When focused laser is used to excite the sample and the prompt
luminescence is measured in the vicinity of the laser spot, a ring, called the
internal ring, formed. While a second ring of PL appears as distant as 1
$\mathrm{mm}$ away from the source, called the external ring. (ii) The
intervening region between the internal and external rings are almost dark
except for some localized bright spots. (iii) Periodic bright spots appear in
the external ring.\cite{Butov2002b} The bright spots follow the external ring
either when the excitation spot is moved over the sample, or when the ring
radius is varied with the excited power. (iv) The PL is eventually washed out
when the temperature is increased.

To explain the above features, a charge separated transportation mechanism was
proposed which gives satisfactory results to the formation of the exciton ring
and the dark region between the internal ring and external ring, and even the
ring in a single quantum well.\cite{butov:117404,rapaport:117405} However, the
physical origin for the periodic bright spots in the external ring is still
controversial. It is commonly believed that the periodic bright spots are
formed due to some kind of instability. Levitov \emph{et al.} considered that
exciton states are highly degenerate and the instability comes from the
stimulated scattering.\cite{Levitov2004a} Sugakov suggested that the
instability is due to the attractive interaction between the high-density
excitons.\cite{Sugakov2004a} In addition, Yang \textit{et al} proposed that
the ring bright spots result from the interplay between the random potential
and the nonlinear repulsive interaction of the condensed
excitons.\cite{Yang2004a}

More recently, great attention has been attracted by a surprising observation
of trapped excitons in an impurity potential well. In the experiment of Lai
\textit{et al.} \cite{LaiCW2004}, they used a defocused laser to excite CQW in
a large area. Excited electrons and holes are collected by an impurity
potential well. The area of the defocused laser spot is typically larger than
the area of impurity potential well. The defoused laser spot can be applied
either away from the impurity potential well or directly on it. It is found
that the PL pattern is much more concentrated than a Gaussian with a central
intensity dip, exhibiting an annular shape with a darker central region. In
particular, with increasing the laser excitation power, exciton cloud first
contracts and then expands. Even more interestingly, pumping by higher-energy
laser, the dip can turn into a tip at the center of the annular cloud. In
fact, this kind of annular shape pattern can also be found in some localized
bright spots between inner ring and external ring of Butov's
experiment.\cite{Butov2002b}

Neither the mechanism of stimulated scattering \cite{Levitov2004a} nor the
mechanism of pure attractive interaction between excitons \cite{Sugakov2004a}
is able to explain the above remarkable phenomenon. With the mechanism of the
stimulated scattering, the stimulated scattering rate should enhance when the
particle density is increased. Consequently, the annular ring should contract
all the way and no expansion is expected. While the mechanism of the pure
attractive interaction between excitons also has the difficulty in explaining
the expansion of excitons.

In this paper, we propose an alternative model to understand the formation of
the bright spots in the exciton ring. We consider that the exciton system can
be described by a nonlinear Schr\"odinger equation which includes an
attractive two-body and a repulsive three-body interactions. The interplay
between these interactions and the kinetic energy can lead to complex
patterns, which are shown to be in great likeness to the experiments. The
nonlinear Schr\"odinger equation is numerically solved and the corresponding
spatial distribution of excitons not only can explain the ring bright spots
but also can describe the contraction and the expansion phenomenon of the
exciton ring. Some detailed features including a tip occurring at the center
of the annular cloud can also be reproduced. Our model predicts some new
patterns at higher-density regime, which can be tested by future experiments.

It is important to emphasize a key element which supports the
proposed model. As pointed out in
Refs.~\cite{butov:117404,rapaport:117405}, when electrons and holes
are excited by laser, they are hot electrons and hot holes
initially. Since the drift speed of hot electrons are larger than
that of hot holes (electron has a smaller effective mass), electron
and holes are indeed charge-separated. No true exciton is formed at
this stage. Because hot electrons and holes have a small
recombination rate, they can travel a long distance from the laser
spot. After a long-distance travel, hot electrons and holes collide
with lattice and are cooled down. The cooling speed of electrons is
faster than that of holes, and consequently cooled electrons and
cool holes will meet. This kind of charge separated transportation
mechanism has been used to interpret the exciton ring
formation.\cite{butov:117404,rapaport:117405} In the experiment by
Butov \textit{et al.} \cite{Butov2002b}, they meet in the region of
the (external) ring. While in the experiment by Lai \textit{et al.}
\cite{LaiCW2004}, they meet in the impurity potential well. It is
believed that cooled electrons and holes will form excitons. Thus
the elemental particles in the external ring \cite{Butov2002b} and
in the impurity potential well \cite{LaiCW2004} are \emph{excitons}.
At this stage, charges are not separated, but coupled or bound
together. They have long life. The interaction between excitons
results in the nonhomogeneous density distribution which in turn
results in the complex PL patterns. Since particles can only move in
CQW, their movements are basically two-dimensional. PL intensity is
directly proportional to the exciton number. In the following
discussion, we simply take exciton probability density distribution
as PL distribution.

\section{Remarks on interactions and equilibrium}

There are two experimental facts which are important to the understanding of
the pattern formation of the indirect excitons. (1) The distribution of the
excitons is inhomogeneous and (2) the inhomogeneous distribution varies with
the exciton number density. To understand these phenomena, the key may lie in
the interactions between excitons. First, it is very clear that the
interaction between the indirect excitons is neither purely attractive, nor
purely repulsive. If the interaction between the indirect excitons is purely
repulsive, it will drive the excitons towards homogeneous distribution and the
exciton cloud will expand with the increase of the exciton number. At present
there is no experimental signature to show this. On the other hand, if the
interaction is purely attractive, the system is expected to collapse when the
exciton density is greater than a critical value to which there is no enough
kinetic energy to stabilize the exciton cloud. Experimentally the collapse of
an exciton cloud has never been observed. In addition, the case of a purely
repulsive or a purely attractive interaction is also against the experimental
fact that the exciton cloud contracts first and expands later when the laser
power is increased. The existence of the attractive interaction does not mean
that the exciton state is unstable against the formation of metallic
electron-hole droplet because the repulsive interaction may dominate over the
attractive one in that regime.

Some remarks are in order on the existence of the attractive interaction.
Different to the direct excitons in a bulk material or in a single quantum
well, the indirect excitons have same polarization direction since the
indirect excitons are formed by electrons and holes which are spatially
separated in different quantum wells. They are aligned dipoles. The
interaction between the indirect excitons contain the dipole-dipole term and
the van der Waals term.\cite{Sugakov2004a} The van der Waals attraction is
given explicitly by the form $-C_{6}/r^{6}-C_{8}/r^{8}-\cdot\cdot\cdot$. When
the spacing of the indirect excitons is large, the dipole-dipole interaction
dominates, so the interaction is effectively repulsive. However, when the
spacing of the indirect excitons becomes small, the van der Waals attraction
will dominate the dipole-dipole interaction. It is found that the interaction
becomes effectively attractive when the separation between two excitons is
about 3 to 6 exciton radii.\cite{Sugakov2004a} In the current experiment, the
exciton density is about $10^{10}/$ $\mathrm{cm}^{2}$. For this density, the
average distance between the indirect excitons is about 100 $\mathrm{nm}$ and
the exciton Bohr radius $a_{B}$ is about 10 $\sim$ 50 $\mathrm{nm}$
\cite{LaiCW2004} or average distance between excitons is about $2\sim10$
exciton radii. In such a case, it is reasonable to assume that the two-body
interaction is in the attractive regime. In addition, when two indirect
excitons approach to each other, the exchange interaction between electrons
becomes important, which also leads to an attractive interaction. In fact, the
attractive interaction between the excitons has been considered as a possible
candidate to describe the pattern formation observed by
experiments.\cite{Sugakov2004a, levitov2005}

In the dilute limit, it is reasonable to assume that the free-energy density
of the system is given by
\begin{equation}
F=F_{0}+V_{ex}n-g_{1}n^{2}+g_{2}n^{3}+\cdot\cdot\cdot~,
\label{free energy density}%
\end{equation}
in terms of the expansion of exciton density $n$. Here $F_{0}$ is the
free-energy density for hot electrons and hot holes (i.e., the case without
true exciton formed), $V_{ex}$ is an external potential, and $g_{1}$ and
$g_{2}$ are (positive) coupling constants associated with two-body and
three-body interactions. The $-$ ($+$) sign with $g_{1}$ ($g_{2}$) term
denotes the attractive (repulsive) nature of the two-body (three-body)
interaction. As elaborated above, the two-body interaction is believed to be
effectively attractive in the system. Nevertheless inclusion of two-body
interaction term only is not possible to give the system a better description
(see later). If one keeps three-body terms in (\ref{free energy density}) as
well, then it is necessary that the three-body interaction must be repulsive
in order to keep the system stable. It will be shown soon that the
inhomogeneous distribution of the exciton can be due to the competition
between the two-body attraction and the three-body repulsion.

The interaction between excitons may be even more complex. It may include the
coulomb interaction (dipole-dipole interaction, van der Waals interaction) and
exchange interaction. It may also include electron-phonon and coulomb
screening effect. In fact, we use two parameters $g_{1}$ and $g_{2}$ to
describe all the above effects.

Another important issue is wether the excitons are in thermal equilibrium. If
the excitons are all in the ground state, or a complete exciton BEC has been
reached, their distribution must be Gaussian-like. The complex pattern
observed in experiments indicates that this is not the case -- quite a large
portion of excitons are in fact in the excited states. Thus it is important to
have a better knowledge on the energy distributions of the trapped excitons.
The energy distribution in turn involves the (complex) energy relaxation and
recombination processes, which have been studied by several experimental and
theoretical groups.\cite{Benisty1991} For a relaxation process, when the
exciton density is low ($n\ll a_{B}^{2}$), the effects due to the
exciton-exciton and the exciton-carrier scattering can be neglected. In this
case, the relaxation time is mainly determined by the scattering of excitons
off acoustic phonons.\cite{Piermarocchi1996} In particular, at low bath
temperatures $(T_{b}<1$ $\mathrm{K})$, this kind of relaxation rate decreases
dramatically due to the so-called "phonon bottleneck"
effects.\cite{Benisty1991} For the recombination process, because the excitons
in the lowest self-trapped level are in a quantum degenerate state, they are
dominated by the stimulated scattering when the occupation number is more than
a critical value. Strong enhancement of the exciton scattering rate has been
observed in the resonantly excited time-resolved PL experiment
\cite{butov:5608}. Therefore, even though the phonon scattering rate is still
larger than the radiative recombination rate, thermal equilibrium of the
system may not be reached. Essentially the distribution may deviate from the
usual Bose one.\cite{Ivanov1999}

\section{The Model}

To proceed with the above analysis, we use the following effective many-body
Hamiltonian to describe the exciton system
\begin{align}
H  &  =\int d\mathbf{r}\psi^{\dag}\left[  -\frac{\hbar^2\nabla^{2}}{2m^{*}}%
+V_{ex}(\mathbf{r})\right]  \psi\nonumber\\
&  -\frac{g_{1}^{\prime}}{2!}\int d\mathbf{r}\psi^{\dag}\psi^{\dag}\psi
\psi+\frac{g_{2}^{\prime}}{3!}\int d\mathbf{r}\psi^{\dag}\psi^{\dag}\psi
^{\dag}\psi\psi\psi, \label{Hamiltonian}%
\end{align}
where $\psi^{\dagger}(\mathbf{r})$ [$\psi(\mathbf{r})$] denotes the creation
(annihilation) of an exciton at the position \textbf{r}, $m^{*}$ is the
effective mass of the exciton, and $V_{ex}$ is the static external potential.
$g_{1}^{\prime}$ is the coupling constant of two-body attraction, while
$g_{2}^{\prime}$ is the coupling constant of three-body repulsion. In writing
down the above Hamiltonian, the interactions between the excitons are assumed
to be the contacted ones (i.e., the $s$-wave approximation is assumed). Under
the mean-field approximation and neglecting the exciton-pair fields like
$\langle\psi\psi\rangle$ and $\langle\psi^{\dagger}\psi^{\dagger}\rangle$ and
the fields involving any three operators, the mean-field Hamiltonian can be
written as
\begin{align}
H\approx\int d\mathbf{r}\psi^{\dag}\left[  -\frac{\hbar^2\nabla^{2}}{2m^{\ast}%
}+V_{ex}-g_{1}n+g_{2}n^{2}\right]  \psi,
\end{align}
where $n=n(\mathbf{r})\equiv\langle\psi^{\dag}\psi\rangle$ is the local
density of excitons at \textbf{r}, $g_{1}\equiv2g_{1}^{\prime}$, and
$g_{2}\equiv9g_{2}^{\prime}.$ The corresponding static \emph{nonlinear}
Schr\"{o}dinger equation reads
\begin{equation}
-\frac{\hbar^2}{2m^{\ast}}\nabla^{2}\psi_{j}+(V_{ex}-g_{1}n+g_{2}n^{2})\psi
_{j}=E_{j}\psi_{j}, \label{the schrodinger equation}%
\end{equation}
where $\psi_{j}$ and $E_{j}$ are the \textit{j-th} eigenstate and eigenvalue,
respectively. It is assumed that the system is in a \emph{quasi-equilibrium}
state, and the spatial distribution of excitons is given by
\begin{equation}
n(\mathbf{r})=\sum_{j=1}^{\mathcal{N}}\eta_{j}(E_{j})|\psi_{j}(\mathbf{r}%
)|^{2}, \label{n-aver}%
\end{equation}
where $\mathcal{N}$ denotes the total number of energy states that
are {\em trapped} and $\eta_{j}$ is an appropriate probability
function associated with the energy level $E_{j}$. As a further
assumption, we take
\begin{equation}
\eta_{j}\equiv\frac{e^{-\beta E_{j} }}{\sum_{j=1}^{\mathcal{N}}e^{-\beta E_{j}
}}, \label{eta}%
\end{equation}
which has the form of Boltzmann distribution. Here $\beta$ is a parameter used
to describe the exciton distribution. The distribution (\ref{eta}) plays a
central role in the calculation, which is shown to lead to qualitatively good
results in agreement with experiments. Other possibility has been tested, but
none of them work.

Numerically it is convenient to first do the following scaling:
$\psi_j({\bf r})/\sqrt{N}\rightarrow \psi_j({\bf r})$,
$Ng_{1}\rightarrow g_{1}$, and $N^{2}g_{2}\rightarrow g_{2}$, such
that Eq.~(\ref{the schrodinger equation}) remains the same look.
In this case, $n({\bf r})$ becomes the probability density which
satisfies the normalization condition $\int _{S}n({\bf r})dS=1$.
Next, rescale $\psi_j({\bf r})\sigma_{\mathrm{PL}}\rightarrow
\psi_j({\bf r})$ and ${\bf r}/\sigma_{\rm PL}\rightarrow {\bf r}$,
Eq.~(\ref{the schrodinger equation}) then reduces to
\begin{equation}
-\frac{1}{2}\nabla^{2}\psi_{j}+(v_{ex}-a_{1}n+a_{2}n^{2})\psi_{j}%
=\varepsilon_{j}\psi_{j}, \label{the reduced schrodinger equation}%
\end{equation}
where $v_{ex}\equiv V_{ex}/\epsilon$, $a_{1}\equiv g_{1}/\left(
\sigma_{\mathrm{PL}}^{2}\epsilon\right)$, $a_{2}\equiv
g_{2}/\left( \sigma_{\mathrm{PL}}^{4}\epsilon\right)$, and
$\varepsilon_{j}\equiv E_{j}/\epsilon$. Here
$\epsilon\equiv{\hbar^{2}}/{m^{\ast}} \sigma_{\mathrm{PL}}^{2}$
with $\sigma_{\mathrm{PL}}$ being chosen by the root-mean-square
radius of the exciton cloud observed by photoluminescence. With
the above scaling, it is found that
$a_{1}^{2}/a_{2}=g_{1}^{2}/g_{2}\epsilon=g_{1}^{2}{m^{\ast}}
\sigma_{\mathrm{PL}}^{2}/g_{2}\hbar^{2}$, which is a constant for
a particular sample. In the following, we shall use two different
values of $a_{1}^{2}/a_{2}$ for the experiments by Lai \emph{et
al.} \cite{LaiCW2004} and by Butov \emph{et al.}
\cite{Butov2002b}.

In connection with real experiments, three important points should be
clarified. (i) The exciton patterns are fully determined by its self-trapped
interaction. The external potential $V_{ex}$ is not the main cause for complex
exciton patterns. In the experiment by Lai \emph{et al.} \cite{LaiCW2004}, the
role of impurity potential well is to collect the hot particles (electrons and
holes) that form excitons. Thus we include a parabolic potential $v_{ex}$ for
calculations in regards to this experiment. However, in the experiment by
Butov \emph{et al.} \cite{Butov2002b}, the ring distribution of exciton is due
to a charge separated transportation mechanism. It is believed that the size
of exciton ring is much larger than that of the impurity potential. The role
of impurity potential is just leading to inhomogenence. So we will set
$v_{ex}=0$ for calculations in regards to this experiment. (ii) When an
electron and a hole form an exction, it is believed that the kinetic energy is
very low. It means that all the excitons are self-trapped in their
self-trapped potential. The particle with energy over self-trapped potential
enegy is not in the self-trapped well, and thus should not take this kinds of
particles into account. (iii) We consider the excitons distributed initially
as a Gaussian. All self-trapped eigenstates $\psi_{j}$ (i.e. $\varepsilon
_{j}<0$) along with $n(x,y)$ [via Eq.~(\ref{n-aver})] are then calculated self-consistently.


\section{Results and discussions}

\subsection{Excitons in impurity potential well}

\begin{figure}[ptb]
\begin{center}
\includegraphics[width=8cm]{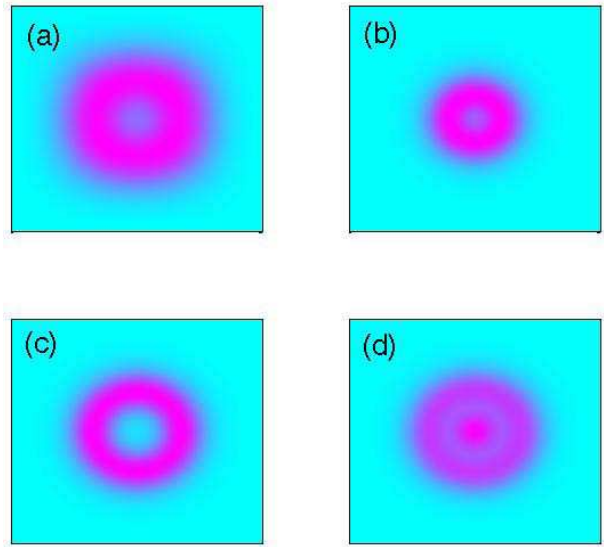}
\end{center}
\caption{Formation of the annular distribution of the self-trapped
excitons as a function of exciton number (a) $a_{1} = 15$, (b)
$a_{1} = 25$, (c) $a_{1} = 40$, (d) $a_{1} = 55$ with $\beta=
0.0001$ in units of $1/\epsilon$ and $a_{2}=0.005a^{2}_{1}$. The
results are intended to be compared with the experiment by Lai
{\em et al.} \cite{LaiCW2004}.}%
\label{fig1}%
\end{figure}

Figure~\ref{fig1} shows the local probability density distribution
$n(x,y)$ of the self-trapped excitons in an impurity potential
well for four different particle number $a_{1}=15$, $25$, $40$,
and $55$. It is natural to quote the particle number $N$ by
$a_{1}$ since $a_{1}\propto g_{1}$ with $g_{1}$ being rescaled to
be $\propto N$. The impurity potential is taken to be parabolic

\begin{equation}
v_{ex}({\bf r}) =\left\{
\begin{array}{cc}
\displaystyle -0.8\left({r/\sigma_{\rm PL}}-1\right)^2 ~~ &
{\rm for}~~r\leq \sigma_{\rm PL}, \\
0 & {\rm otherwise},
\end{array}
\right.  \label{V}
\end{equation}
with a cutoff in simulating the real system. The plots are on a 2D
plane and the color scale denotes the relative amplitude of the
local density $n(x,y)$.  When the irradiating laser power is low
(therefore, the self-trapped particle number is few), the dilute
exciton cloud is diffused because the attractive interaction
($g_{1}$) is weak. When the laser power is increased (therefore,
$g_{1}$ increases), stronger and stronger attraction drives the
exciton cloud to shrink. When the particle number is further
increased, the repulsive interaction ($g_{2}$) becomes more
important, and eventually dominates over the attractive interaction.
As a consequence, the exciton cloud expands again. The evolution of
the exciton cloud with the exciton number (the laser power) is in
good agreement with the experimental observation by Lai $et~al.$
\cite{LaiCW2004}. 
\begin{figure}[ptb]
\begin{center}
\includegraphics[width=8cm]{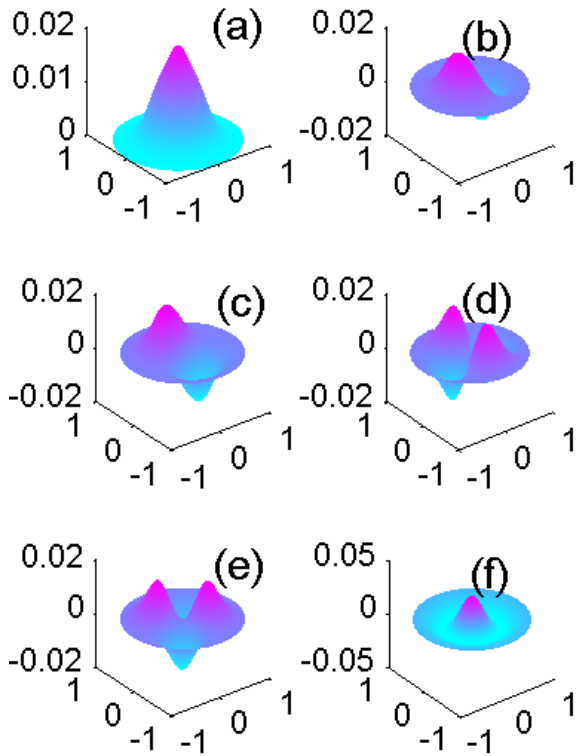}
\end{center}
\caption{The corresponding lower-energy wave functions $z=\psi_{i}(x,y)$
related to the case of Fig.~\ref{fig1}(d).}%
\label{fig2}%
\end{figure}

For a better illustration, the corresponding wave functions of the discrete
energy levels related to Fig.~\ref{fig1}(d) are plotted in Fig.~\ref{fig2}.
The wave functions related the distributions in Fig.~\ref{fig1}(a)-(c) are
similar to those in Fig.~\ref{fig2}.
When the particle number is few [Fig.~\ref{fig1}(a)], the number of the
self-trapped energy eigenstates involved is fewer. The probability density
distribution $n$ is mainly determined by the superposition of the ground state
and the first excited states. The ground-state wave function is $s$-wave with
a peak at the center [see wave function in Fig.~\ref{fig2}(a)], while the
first excited states are nearly twofold degenerate $p$ wave with a node at the
center [see wave functions in Fig.~\ref{fig2}(b) and (c)]. Superposition of
these two states then lead to an annular distribution with a dip in the
center. With increasing of the particle number [Fig.~\ref{fig1}(b) and (c)],
the number of the self-trapped energy eigenstates involved increases and the
second excited states start to intervene. Second excited states are twofold
degenerate $d$ waves [see Fig.~\ref{fig2}(d) and (e)]. Thus a ring with much
higher contrast is obtained. Further increasing the particle number, the third
excited state [see Fig.~\ref{fig2}(f)] then starts to intervene the system.
The superposition results an annual distribution with a tip at the center.

It is useful to estimate the values of $g_{1}$ and $g_{2}$ with
respect to the real system. Taking Fig.~\ref{fig1}(b) as an
example, it is estimated that $n\approx3.0\times10^{10}$
cm$^{-2}$. Since the experiment gives the root-mean-square radius
$\sigma_{\mathrm{PL}}= 10$ $\mu$m, the trapped exciton number
$N=\pi \sigma_{\mathrm{PL}}^{2}n\approx9.4\times10^{4}$. One then
obtains
\begin{equation}
g_{1}=\frac{\hbar^{2}a_{1}}{m^{\ast}\sigma_{PL}^{2}N}\times\sigma_{PL}%
^{2}\approx4.\,\allowbreak93\times10^{-20}\text{meV}%
\end{equation}
and%
\begin{equation}
g_{2}=\frac{\hbar^{2}a_{2}}{m^{\ast}\sigma_{PL}^{2}N^{2}}\times\sigma_{PL}%
^{4}\approx6.\,\allowbreak54\times10^{-36}\text{meV.}%
\end{equation}
Besides, if one assumes that the attractive interaction comes from an $s$-wave
scattering, then the $s$-wave scattering length%
\begin{equation}
a=\frac{a_{1}}{4\pi N}\approx2.\,\allowbreak1\times10^{4}\text{ nm.}%
\end{equation}
Moreover, since the exciton Bohr radius $a_{B}=10\sim50$ nm, the ratio
$a/a_{B}=400\sim2000$. If we assume $\beta\equiv1/k_{B}T$, the exciton
temperature
\begin{equation}
T=\frac{{\hbar^{2}}}{{m^{\ast}}\sigma_{\mathrm{PL}}^{2}{k}_{B}\beta}%
\approx0.2\text{K.}%
\end{equation}

\subsection{Fragmented exciton ring}

\begin{figure}[h]
\begin{center}
\includegraphics[width=8cm]{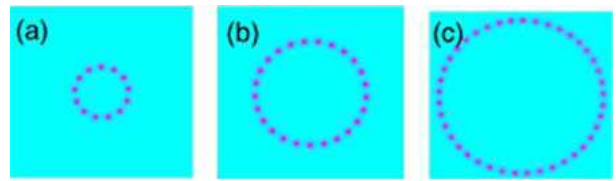}
\end{center}
\caption{The formation of the bright ring spots for different
exciton numbers: (a) $a_{1} = 80$, (b) $a_{1} = 150$, and (c)
$a_{1} = 250$. $\beta= 10^{-5}$ and $a_{2}=10^{-5}a^{2}_{1}$. The
results are intended to be compared with the experiment by Butov
{\em et al.} \cite{Butov2002b}.}%
\label{fig3}%
\end{figure}

The same physical picture can also be employed to explain the ring
bright spots observed by Butov $et~al.$ \cite{Butov2002b} In the
present case, the exciton ring has formed with the charge
separated mechanism.\cite{butov:117404, rapaport:117405} The
initial distribution of excitons is assumed to be homogenous in
the ring. Considering that $\sigma_{\mathrm{PL}}\sim50$$\mu m$,
which is about 5 times larger than that in Fig.~\ref{fig1}, so the
parameter $a_{1}$ is approximately 25 times larger than the
impurity potential well case. Figs.~\ref{fig3}(a) - \ref{fig3}(c)
show the formation of the bright ring spots for different exciton
numbers and the ring radii (the exciton density was kept as a
constant). With increasing the exciton number, the ring radii (so
as the number of the spots) increases, but the density of bright
spot remains unchanged. The periodic spots and the change of the
spots number with the exciton number (or the laser power) are in
qualitative consistence with the experimental
observations.\cite{Butov2002b} In a real sample, non-homogeneity
and some impurity may exist in the system. It is believed that the
distorted and nonhomogeneous patterns result due to the impurity
potential. To give a better fitting to the experiments, the
external potential $V_{ex}$ should be restored in the
Schr\"{o}dinger equation (\ref{the schrodinger equation}).

According to our picture, the physical origin of the bright ring spots is the
consequence of the competition between the two-body attractive and three-body
repulsive interactions and the kinetic energy. If excitons are uniformly
distributed on the external ring initially, the attractive interaction will
drive the excitons to move together. When the local density reaches one
certain value, kinetic energy will drive the high-density excitons to diffuse.
At the same time, the repulsive interaction also hinders further increasing of
the exciton density. The competitive consequence leads to form a series of
clusters on the external ring. The size of these cluster is determined by the
ratio of these three effects.

\subsection{High-density patterns}

\begin{figure}[h]
\begin{center}
\includegraphics[width=8cm]{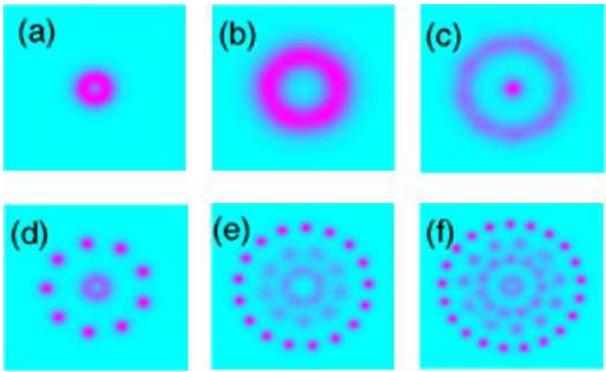}
\end{center}
\caption{The density distribution for different exciton numbers: $a_{1}=$ (a)
$20$, (b) $30$, (c) $60$, (d) $80$, (e) $200$, (f) $300$. Other parameters are
the same as those in Fig.~\ref{fig2}.}%
\label{fig4}%
\end{figure}

The above calculations indicate that the phenomenological
Schr\"{o}dinger equation (\ref{the schrodinger equation}) gives a
good description to the exciton distribution, both in an impurity
potential well and in the external ring. One open question is what
will happen when the particle number in an impurity potential well
is much increased. Fig.~\ref{fig4} shows the numerical results for
various exciton densities. When the exciton density is low
[Fig.~\ref{fig4}(a)], an annular distribution is observed. With
increasing the density of the excitons, the annular exciton cloud
expands and a tip emerges at its center [Figs.~\ref{fig4}(b) \&
(c)]. This phenomena has been observed in the experiment by Lai
\emph{et al.}.\cite{LaiCW2004} When the density is increased
further, the annular exciton cloud is unstable and a series of
clusters form [Fig.~\ref{fig4}(d)]. In the case that the density is
high enough, the competition between different interactions can in
fact lead to much complex patterns beyond the present experimental
observations. The clusters eventually form a pattern of triangle
lattice [Figs.~\ref{fig4}(e) \& (f)]. We point out that all these
patterns could be observable under current experimental conditions
when the density is large enough

\section{Conclusion}

Finally, some remarks are in order on the temperature effect. When
the bath temperature is low, excitons are cooled and have relatively
low momenta. The self-trapped interaction can confine most of
excitons. However, in the low momentum case, cooling efficiency is
low while luminous efficiency is high, excitons can not reach the
thermal equilibrium state. Due to the competition between the
self-trapped and kinetic energies, complex exciton patterns occur
(as discussed above). When the temperature is increased, excitons
are not fully cooled and correspondingly self-trapped interaction
confines only part of the excitons. The attractive interaction can
not compensate the exciton kinetic energy and excitons will
distribute homogenously in 2D plane. In this case, the pattern is
washed out. If the temperature is higher than the indirect exciton
binding energy $\sim3.5$ $\mathrm{meV},$
\cite{Snoke2004b,szymanska:193305} most of excitons become ionized
and are in a plasma state. No pattern can be observed in this case.
In order to realize exciton BEC, the further experimental work
should look for effective method to obtain the excitons with low
combination rate and short relaxation time. The further theoretical
work should focused on the exciton interaction beyond the mean field
approximation.

To summarize, we have demonstrated that the exciton distributions observed in
experiments can be explained by the competition between the self-trapped
interaction and the kinetic energy. A nonlinear Schr\"odinger equation
including short-range two-body attractive and three-body repulsive
interactions is used to describe the exciton behavior. The interplay among the
two-body interaction, the three-body interaction, and the kinetic energy not
only explains the experimental observations, but also leads to rich patterns,
which could be tested in future experiments.

\begin{acknowledgments}
We acknowledge fruitful discussions with T. Xiang, L. Yu, Z. B. Su, G. H. Ji
and J. H. Yuan. This work was supported by the National Natural Science
Foundation of China (Grant No. 10347149), National Basic Research Program of
China (Grant No. 2005CB32170X), and National Science Council of Taiwan (Grant
No. 93-2112-M-003-015).
\end{acknowledgments}


\end{document}